\documentclass[aps,pra,twocolumn,notitlepage,superscriptaddress,showpacs,10pt,longbibliography]{revtex4-2} 

\usepackage{amsmath}    
\usepackage{amsfonts}
\usepackage{bm}
\usepackage{amssymb}
\usepackage{appendix}
\usepackage{graphicx}   
\usepackage{xcolor}      
\usepackage{subfigure}  
\usepackage{comment}
\usepackage{siunitx}
\usepackage{cancel}
\usepackage{enumitem}

\usepackage{soul}
\usepackage[normalem]{ulem}

\mathchardef\mhyphen="2D

\renewcommand{\vec}[1]{\mathbf{#1}}

\begin{document}

\title{Observation of bound states in the continuum embedded in symmetry bandgaps} 
\author{Alexander Cerjan}
\thanks{These two authors contributed equally}
\affiliation{Department of Physics, The Pennsylvania State University, University Park, Pennsylvania 16802, USA}
\affiliation{Sandia National Laboratories, Albuquerque, New Mexico 87185, USA}
\affiliation{Center for Integrated Nanotechnologies, Sandia National Laboratories, Albuquerque, New Mexico 87185, USA}
\author{Christina J{\" o}rg}
\thanks{These two authors contributed equally}
\affiliation{Department of Physics, The Pennsylvania State University, University Park, Pennsylvania 16802, USA}
\author{Sachin Vaidya}
\affiliation{Department of Physics, The Pennsylvania State University, University Park, Pennsylvania 16802, USA}
\author{Shyam Augustine}
\affiliation{Physics Department and Research Center OPTIMAS, University of Kaiserslautern, D-67663 Kaiserslautern, Germany}
\author{Wladimir A.\ Benalcazar}
\affiliation{Department of Physics, The Pennsylvania State University, University Park, Pennsylvania 16802, USA}
\author{Chia Wei Hsu}
\affiliation{Ming Hsieh Department of Electrical Engineering, University of Southern California, Los Angeles, California 90089, USA}
\author{Georg von Freymann}
\affiliation{Physics Department and Research Center OPTIMAS, University of Kaiserslautern, D-67663 Kaiserslautern, Germany}
\affiliation{Fraunhofer Institute for Industrial Mathematics ITWM, 67663, Kaiserslautern, Germany}
\author{Mikael C. Rechtsman}
\affiliation{Department of Physics, The Pennsylvania State University, University Park, Pennsylvania 16802, USA}

\date{\today}

\begin{abstract}
    In the last decade, symmetry-protected bound states in the continuum (BICs) have proven to be an important design principle for creating and enhancing devices reliant upon states with high quality ($Q$) factors, such as sensors, lasers, and those for harmonic generation. However, as we show, current implementations of symmetry-protected BICs in photonic crystal slabs can only be found at the center of the Brillouin zone and below the Bragg-diffraction limit, which fundamentally restricts their use to single-frequency applications. By 3D-micro printing a photonic crystal structure using two-photon polymerization, we demonstrate that this limitation can be overcome by altering the radiative environment surrounding the slab to be a three-dimensional photonic crystal. This allows for the protection of a line of BICs by embedding it in a symmetry bandgap of the crystal. Moreover, we experimentally verify that just a single layer of this photonic crystal environment is sufficient. This concept significantly expands the design freedom available for developing next-generation devices with high-$Q$ states.
\end{abstract}

\maketitle

\section{Introduction}
Over the last decade, bound states in the continuum (BICs) have emerged as an important design principle for creating systems with high quality-factor ($Q$) states to enhance light-matter interactions. BICs are states with theoretically infinite lifetimes despite the availability of a radiative continuum at the same frequency \cite{hsu_bound_2016}. By operating a system in the vicinity of a BIC in some generalized parameter space, arbitrarily large quality factors can be realized that allow the $Q$ of the system to be tailored to the specific needs of the device. Although several different mechanisms can be used to create BICs in optical systems \cite{friedrich_interfering_1985,paddon_two-dimensional_2000,ochiai_dispersion_2001,pacradouni_photonic_2000,fan_analysis_2002,bulgakov_bound_2008,plotnik_experimental_2011,hsu_observation_2013,gomis-bresco_anisotropy-induced_2017,azzam_formation_2018,mukherjee_topological_2018,doeleman_experimental_2018,zhang_observation_2018,cerjan_bound_2019,jin_topologically_2019,kim_optical_2019,overvig_selection_2020,yin_observation_2020,benalcazar_bound_2020,cerjan_observation_2020,hayran_capturing_2021,kang_merging_2021,vaidya_point-defect_2021}, the preponderance of interest has focused on using the structure's symmetry to protect a BIC from radiating for two reasons: First, symmetry-protection is predictive, and planar systems possessing $180^\circ$ rotational symmetry about the $z$-axis ($C_2$) will generically possess BICs. Second, no fine tuning of the structure is necessary to adjust the BIC's location in wavevector space, since the BIC is guaranteed to exist at normal incidence. As such, symmetry-protected BICs in photonic crystal slabs and all-dielectric metasurfaces have been demonstrated to enable or enhance a wide variety of different applications, such as sensors \cite{yanik_seeing_2011,zhen_enabling_2013,romano_label-free_2018}, high power on-chip lasers \cite{hirose_watt-class_2014,kodigala_lasing_2017}, vortex lasers \cite{wang_generating_2020,huang_ultrafast_2020}, harmonic generation \cite{minkov_zero-index_2018,minkov_doubly_2019,liu_high-q_2019,koshelev_subwavelength_2020,wang_doubly_2020,ginsberg_enhanced_2020}, and increased control over transmission and reflection spectra \cite{campione_broken_2016,koshelev_asymmetric_2018,koshelev_meta-optics_2019,zito_observation_2019,gorkunov_metasurfaces_2020}.

Despite these successes of using symmetry protection to create BICs for next-generation devices, current designs for achieving symmetry-protected BICs in slab geometries are fundamentally limited. As we rigorously prove in this work, if one is restricted to engineering structures within the slab, symmetry-protected BICs always appear as isolated states that can \textit{only} exist at normal incidence, below the Bragg-diffraction limit. Thus, any application requiring a range of high-$Q$ states, such as operating a sensor at multiple frequencies simultaneously, or steering a laser beam using lasing modes in the same device with different in-plane wavevectors \cite{kurosaka_-chip_2010}, cannot be realized with existing system designs using symmetry-protected BICs.

We show that such limitations of two-dimensional BICs can be overcome by designing the environment surrounding the slab, as opposed to the slab itself. We theoretically propose and experimentally realize a line of BICs in a ``symmetry bandgap'' by embedding a homogeneous slab in a three-dimensional rectangular woodpile photonic crystal environment. 
Here, we define a symmetry bandgap as a wavevector-dependent frequency range along a high-symmetry line (or at a high-symmetry point) in which only a subset of the possible symmetry representations are present among states in the environment. Thus, a slab state with the appropriate symmetry representation inside a symmetry bandgap of the surrounding radiative environment is necessarily a symmetry-protected BIC.
Our monolithic structure is fabricated entirely from photoresist polymer using two-photon polymerization \cite{von_freymann_three-dimensional_2010,hohmann_three-dimensional_2015} and characterized using angle-resolved Fourier-transform infrared (FTIR) spectroscopy \cite{Li:06}. The line of symmetry-protected BICs is directly observed as the vanishing linewidth of a resonance of the slab along a high-symmetry line of the system. 

\section{Results}
\subsection{Symmetry restrictions for finding BICs}
\begin{figure}[t]
    \centering
    \includegraphics[width=1.0\columnwidth]{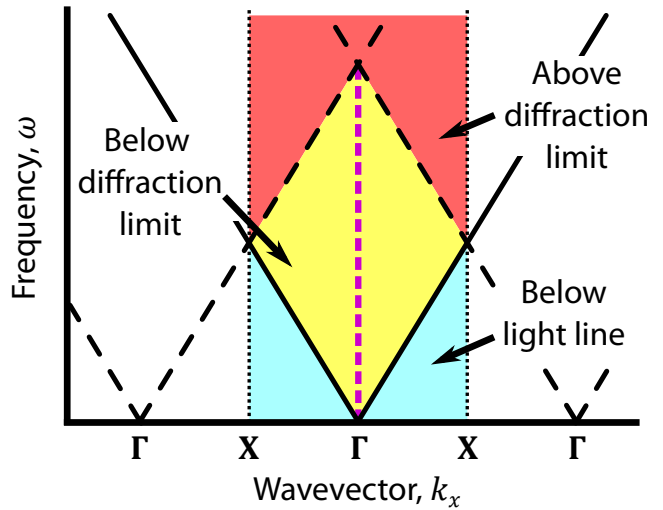}
    \caption{ \small{
        \textbf{Radiative channel limits in the Brillouin zone for homogeneous, isotropic environments.}
        Schematic showing the regions of the Brillouin zone in the $k_y = 0$ plane which are below the light line (cyan), below the first Bragg-diffraction limit (yellow), and above the first Bragg-diffraction limit (red). High symmetry points are marked, $\bold{\Gamma} = (0,0)$, and $\textbf{X} = (\pi/a, 0)$. For environments which are homogeneous and isotropic, symmetry-protected BICs can only exist at $\bold{\Gamma}$ and below the Bragg-diffraction limit, indicated as the magenta dashed line. Only the first Brillouin zone is shaded as a reminder that slab resonances are only uniquely defined in $\vec{k}_\parallel$ up to a reciprocal lattice vector.
      \label{fig:0}}}
\end{figure}

The capacity for a device to support symmetry-protected BICs can be viewed from the perspective of the environment: For a symmetry-protected BIC to exist, at least one symmetry representation of the device must be absent in the available radiative channels of the surrounding environment. As an example, consider the radiation from a resonance of a photonic crystal slab with in-plane wavevector $\vec{k}_\parallel = \bold{\Gamma} = (0,0)$ into the surrounding air. At low frequencies, $\omega$, conservation of in-plane momentum dictates that the only available radiative channels above the slab that the resonance can couple to are $s$- and $p$-polarized outgoing plane waves with $\vec{k} = (0,0,\omega/c)$ (yellow region in Fig.\ \ref{fig:0}). Both of these channels are odd with respect to rotation about the $z$-axis by $180^\circ$ ($C_2$), as this rotation leaves $\vec{k}$ invariant, but reverses the direction of the polarization. The same is true of the two available radiative channels below the slab. Thus, any photonic crystal slab that is $C_2$ symmetric and surrounded by air will generically possess states at $\bold{\Gamma}$ that are even with respect to $C_2$ and cannot radiate due to this symmetry mismatch, i.e.\ these states are necessarily symmetry-protected BICs. In contrast, away from $\bold{\Gamma}$, but still in the interior of the first Brillouin zone, degenerate $s$- and $p$-polarized plane waves span all possible in-plane symmetry representations, so all resonances of the slab will generally radiate at these $\vec{k}_\parallel$, which prohibits the formation of symmetry-protected BICs, see Supplemental Materials.

In an isotropic and homogeneous environment, additional radiative channels become available above the Bragg-diffraction limit \cite{joannopoulos}, $n \omega/c > |\vec{k}_\parallel \pm \vec{b}_{i}|$ (red region in Fig.\ \ref{fig:0}). Here, $i = 1,2$, $\vec{b}_{1,2}$ are the reciprocal lattice vectors of the slab, and $n$ is the environment's refractive index. These extra channels correspond to light acquiring additional momentum from the periodicity of the lattice as it radiates. Note, along the first Brillouin zone's boundary, frequencies which are above the light line, $n \omega/c > |\vec{k}_\parallel|$, are necessarily above the Bragg-diffraction limit. For such an environment, one can prove that these additional channels span all of the in-plane symmetries of the system for every $\vec{k}_\parallel$, prohibiting the formation of symmetry-protected BICs above the Bragg-diffraction limit.

To give an abbreviated argument for this statement, consider a slab resonance with frequency $\omega$ and in-plane wavevector $\vec{k}_\parallel \ne \bold{\Gamma}$, and an available radiative channel with wavevector $\vec{k} = (\vec{k}_\parallel, k_z=(n^2\omega^2/c^2 - |\vec{k}_\parallel|^2)^{(1/2)})$. Assume $\vec{k}_\parallel$ is a high-symmetry point where it may be possible for the slab resonance to possess a symmetry representation that does not exist in the environment, and let $S$ be an in-plane symmetry operation which leaves $S\vec{k}_\parallel$ equivalent to $\vec{k}_\parallel$, i.e.\ $S\vec{k}_\parallel = \vec{k}_\parallel + \sum_{i=1,2} m_i \vec{b}_i$ for $m_i \in \mathbb{Z}$. Then there are two possibilities: 
\begin{enumerate}[noitemsep,nolistsep,label=(\arabic*)]
    \item If $S\vec{k} = \vec{k}$, then for all 17 of the 2D space groups $S$ is a reflection or glide operation, which the resonance can be even or odd with respect to, but both of these possibilities are spanned by degenerate $s$- and $p$-polarized plane waves, as the two polarizations behave oppositely under reflections and glides.
    \item If $S\vec{k} \ne \vec{k}$, then these wavevectors correspond to orthogonal radiating plane waves, linear combinations of which necessarily span the possible in-plane symmetry representations of $S$.
\end{enumerate}
Together, these two statements show that there will always be an available radiative channel for any slab resonance to couple to above the Bragg-diffraction limit in a homogeneous and isotropic environment, precluding any symmetry-protected BICs from appearing. A rigorous proof using representation theory can be found in the Supplementary Materials.

In sum, symmetry-protected BICs in periodic planar structures surrounded by homogeneous, isotropic environments can only be found at $\vec{k}_\parallel = \bold{\Gamma}$ and below the Bragg-diffraction limit ($n \omega/c < |\vec{b}_{1,2}|$ for this wavevector) --- no amount of engineering of the slab can overcome this limitation. 

\begin{figure}[t]
    \centering
    \includegraphics[width=1.0\linewidth]{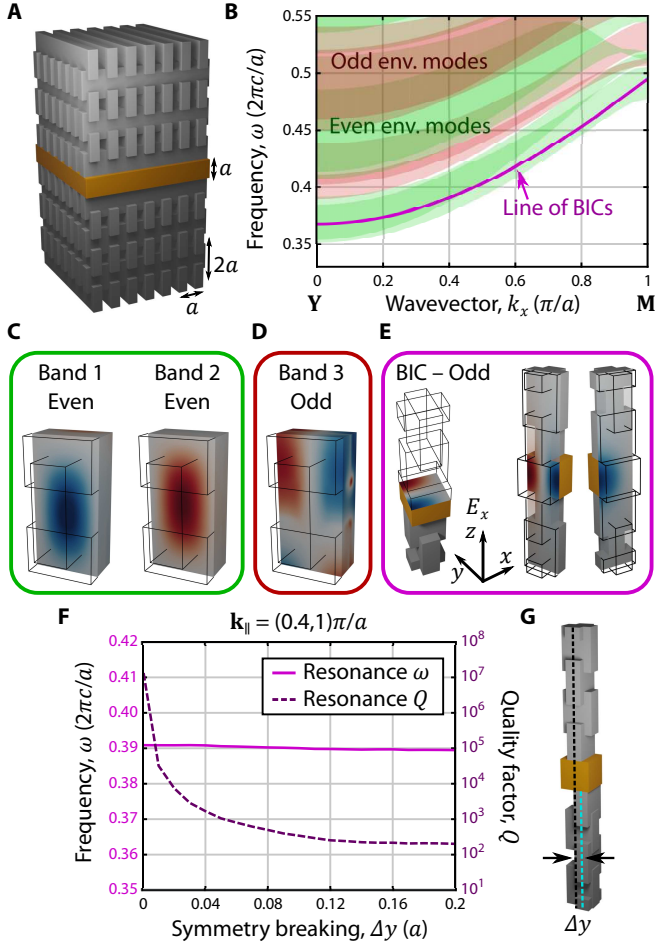}
    \caption{ \small{
        \textbf{Symmetry-protected BIC in a slab embedded in a rectangular woodpile photonic crystal.}
        (a) Schematic of a solid slab (orange) embedded in a rectangular woodpile photonic crystal environment (gray). (b) Projected-in-$k_z$ band structure of the woodpile environment along the $\bold{Y}$--$\bold{M}$ high-symmetry line. Green (red) regions indicate where a projected band of the woodpile are even (odd) with respect to reflection about the $xz$-plane. A line of BICs of the slab which are odd about this reflection symmetry is shown in purple. (c,d,e) $E_x$ modal profiles of the first three bands of the environment and one of the BICs at $\vec{k}_\parallel a / \pi = (0.4, 1)$, which are even (c) or odd (d,e) with respect to this reflection symmetry. (f) Quality factor and resonance frequency of the BIC at $\vec{k}_\parallel a / \pi = (0.4, 1)$ as the upper and lower woodpile environments are displaced by $\Delta y$, sketched in (g).
      \label{fig:1}}}
\end{figure}

\begin{figure*}[tb]
    \centering
      \includegraphics[width=1.0\linewidth]{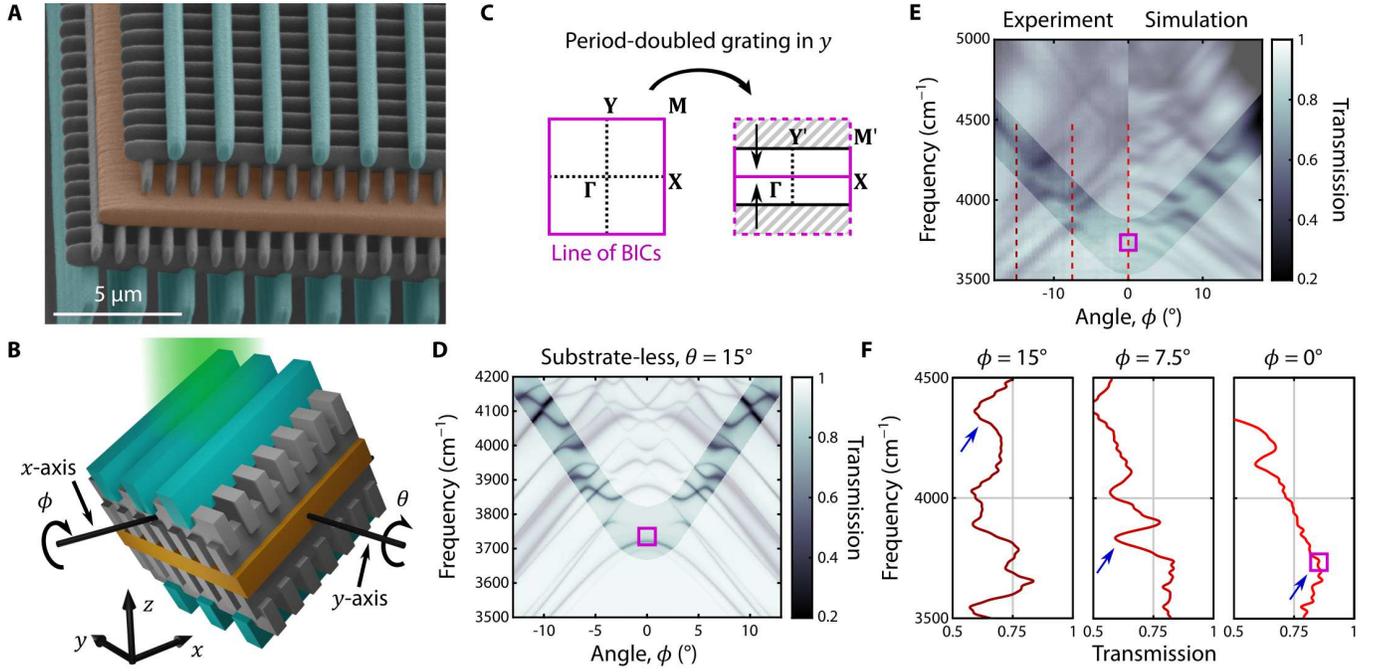}
    \caption{ \small{ 
        \textbf{Observation of a symmetry-protected BIC away from $\bold{\Gamma}$.}
        (a) False color scanning electron microscope image of a photonic crystal slab (orange) embedded in a rectangular woodpile photonic crystal environment (grey) with period-doubled grating (cyan). (b) Sketch of the structure with angles $\theta$ and $\phi$ relevant for the measurements. (c) Schematic showing the folding of the first Brillouin zone due to the period-doubled grating. (d) Simulated transmission spectra of the substrate-less structure along $\phi$. (e) Transmission spectra of the BIC structure (left experiment, right simulation) along $\phi$. In (d,e) shading is used to highlight the resonance which becomes a BIC at $\phi = 0$. For (d,e), $\theta = 15^\circ$ (which corresponds to $k_x = 0.19(\pi/a)$), and the purple square indicates the frequency of the BIC calculated for a slab surrounded by an infinite woodpile environment. Slices along the red dashed lines are depicted in (f), where we can see resonance features in the transmission for $\phi=15^{\circ}$ and $\phi=7.5^{\circ}$, that vanish at the BIC. Blue arrows are a visual guide to the resonance in question.
      \label{fig:2}}}
\end{figure*}

\subsection{Realizing BICs through environmental design}

To realize symmetry-protected BICs anywhere else in the Brillouin zone of a planar structure, one is forced to either break isotropy or homogeneity of the radiative environment. Breaking isotropy (for example, by using a birefringent environment) lifts the degeneracy between $s$- and $p$-polarized plane waves, allowing for the possibility of a symmetry bandgap depending on the orientation of the underlying material's crystalline axes \cite{gomis-bresco_anisotropy-induced_2017,mukherjee_topological_2018}.

Here we break the homogeneity of the environment surrounding the slab by using a three-dimensional photonic crystal as the radiative environment. In addition to breaking the degeneracy between $s$- and $p$-polarized plane waves \cite{cerjan_bound_2019}, this also allows for radiative channels with $S\vec{k} \ne \vec{k}$ to still be equivalent and correspond to the same radiative channel, as the environment now has discrete, not continuous, translational symmetry. As such, a three-dimensional photonic crystal environment can possess symmetry bandgaps for $\vec{k}_\parallel \ne \bold{\Gamma}$.

It may seem counter-intuitive that by reducing the symmetry of the environment from continuous to discrete translational symmetry, one somehow gains access to additional forms of symmetry protection in the slab, but patterning the environment provides two fundamental benefits beyond breaking the degeneracy between the two polarizations. First, using a three-dimensional photonic crystal environment with discrete translational symmetry in $z$ means that radiative channels have maximum frequency cutoffs, which is not the case in free space or homogeneous birefringent environments, and allows for symmetry bandgaps to appear even in high-frequency regions of the environment's band structure. Second, in-plane discrete translational symmetries allow for the appearance of additional high-symmetry lines in the environment's band structure along the boundary of the first Brillouin zone; there is nothing remarkable about the $\bold{Y}$--$\bold{M}$ line in free space, but there is in the rectangular woodpile environment.

In particular, we embed a homogeneous slab inside a three-dimensional rectangular woodpile photonic crystal environment, Fig.\ \ref{fig:1}a. This changes the radiative channels of the environment to be the projected-in-$k_z$ bands of the photonic crystal environment, as the presence of the slab breaks translational symmetry in $z$. As the mostly-evanescent tails of the slab's resonances overlap with the periodic environment, no patterning of the slab is necessary for its resonances to form a photonic crystal slab band structure. Moreover, as the slab is homogeneous, or more generally so long as the slab and the environment have commensurate in-plane lattice vectors, $\vec{k}_\parallel$ is conserved during radiation. For this photonic crystal environment, we find that along the $\bold{Y}$--$\bold{M}$ ($\bold{X}$--$\bold{M}$) high-symmetry line, where the system is reflection symmetric about the $xz$-plane ($yz$-plane), the two lowest frequency radiative channels are both even with respect to this symmetry, Figs.\ \ref{fig:1}b-d. As such, the rectangular woodpile exhibits a symmetry bandgap in its two lowest-frequency bands along these high-symmetry lines.

Within this symmetry bandgap, states of the photonic slab with the opposite symmetry are necessarily symmetry-protected BICs, such as the state shown in Fig.\ \ref{fig:1}b,e. Moreover, we can confirm that the exponential confinement of the state is due to its mismatched symmetry by displacing the rectangular woodpile environments above and below the slab to break the $xz$-plane reflection symmetry of the whole structure. This displacement yields environments above and below the slab that no longer share the same reflection plane, so that their radiative channels necessarily span all of the possible symmetries along the $\bold{Y}$--$\bold{M}$ high-symmetry line, i.e., there is no longer a symmetry bandgap. As such, this perturbation significantly decreases the $Q$ of the slab resonance, as shown in Fig.\ \ref{fig:1}f,g.

\subsection{Experimental observation}
We can experimentally observe this line of symmetry-protected BICs using only a single layer of photonic crystal environment. The entire system is fabricated by direct laser writing on top of a glass substrate using two-photon polymerization of a low-index photoresist, $\varepsilon = 2.34$, shown in Fig.\ \ref{fig:2}a (for more details see the Methods section). The rods comprising the rectangular woodpile have a cross-section with width $a/2$ and height $1.4a$, and each layer of the woodpile is separated vertically in $z$ by $a$ (layers overlap), where $a = \SI{1}{\micro\meter}$ is the lattice constant in $x$ and $y$. The slab has height $a$. Simulations show that due to the strong localization of the BIC slab modes, only a single unit cell of the woodpile environment is required on each side of the slab to preserve the exponential localization of the resonance due to symmetry, see Supplemental Materials. Additionally, as the line of BICs in the setup of Fig. \ref{fig:1} is under the light line of free space, we use a period-doubled grating written on the top and bottom of the structure (cyan elements in Fig.\ \ref{fig:2}a,b), for in- and out-coupling. This remaps the line of BICs to be along $\bold{\Gamma}$--$\bold{X}$, and thus above the light line of free space, see Fig.\ \ref{fig:2}c. The grating elements have widths $0.8a$, and heights $1.4a$ (above) and $4a$ (below), see Methods. Together, the combined effects of the truncated environment and grating yield a $Q$-factor of the confined resonance in excess of $10^6$, see Supplemental Materials. Transmission spectra are taken with a Fourier-transform infrared spectrometer by tilting the samples by the angle $\theta$ about the $y$-axis, while sweeping through the angle $\phi$ which rotates the sample about the $x$-axis, see Fig.\ \ref{fig:2}b. In this measurement scheme, $k_x = (\omega / c) \sin(\theta)$, $k_y = (\omega / c) \cos(\theta) \sin(\phi)$, and the $\bold{\Gamma}$--$\bold{X}$ line corresponds to $\phi = 0$.

\begin{figure}[t]
    \centering
    \includegraphics[width=1.0\linewidth]{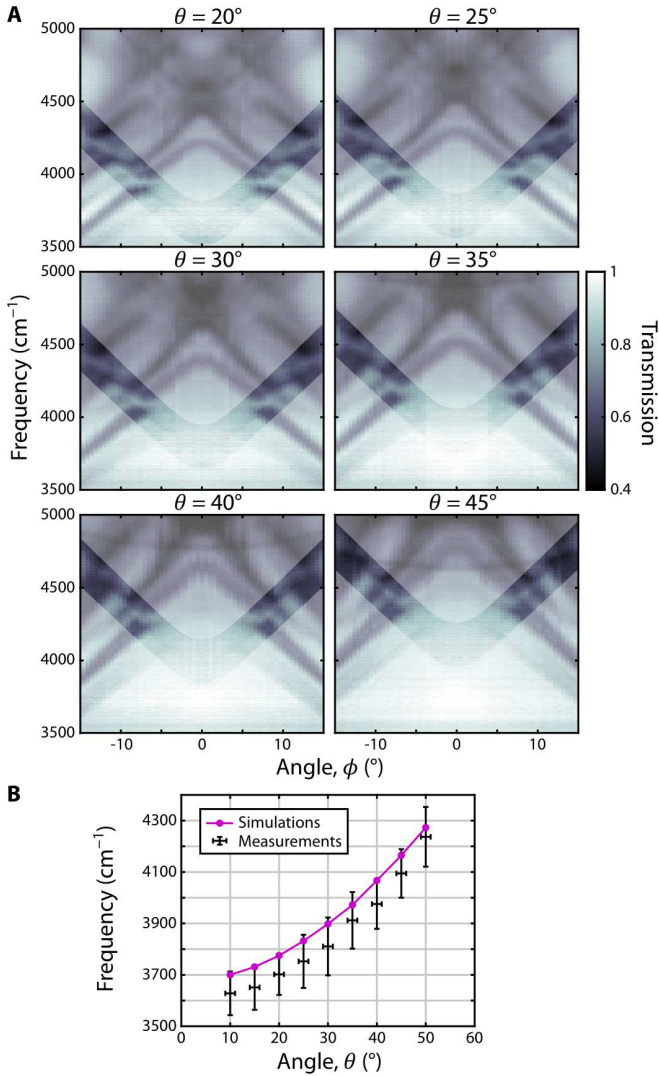}
    \caption{ \small{
        \textbf{Line of symmetry-protected BICs.}
        (a) Experimentally observed transmission spectra of the BIC structure as $\phi$ is varied and for different positions along $\bold{\Gamma}$--$\bold{X}$, given by $\theta$. Shading is used to highlight the resonance which vanishes around $\phi = 0$. (b) BIC frequencies as extracted from the measurements (black) and the simulations (magenta) versus $\theta$. The plotted values of increasing $\theta$ correspond to $k_x = (\pi/a) [0.13, 0.19, 0.26, 0.32, 0.39, 0.46, 0.52, 0.59, 0.66]$.
      \label{fig:3}}}
\end{figure}

The resonance of the slab which becomes a symmetry-protected BIC at $\phi=0$ can be identified in the angle-resolved transmission spectrum as a concave-up series of avoided crossings for $|\phi|>0$, whose linewidth approaches zero as $|\phi|\rightarrow 0$. This process is shown in simulations of a substrate-free system in Fig.\ \ref{fig:2}d, where the linewidth of the resonance becomes too narrow to be resolved at this scale for $|\phi| \le 3^\circ$. Figure \ref{fig:2}e shows a comparison of the experimental observation with simulation of the complete fabricated structure where again the resonance becomes too narrow to be resolved for $|\phi| \le 6^\circ$. In the simulation results shown in Fig.\ \ref{fig:2}e, only the specular transmission has been retained and this channel has been averaged over $\theta \pm 2^\circ$ and $\phi \pm 0.3^\circ$ to mimic the behavior of the spot shape and pinhole used in the experimental measurements (see Methods). The vanishing linewidth can also be seen in a series of slices of the experimental data, Fig.\ \ref{fig:2}f. The discrepancies seen between the experiment and simulation in Fig.\ \ref{fig:2}e are likely due to the difficulty in determining the exact structure parameters using scanning electron microscopy at the edge of the device. Additionally, the resonance at $\phi = 0$ can be revealed in our experimental observations by purposefully breaking the symmetry of the structure which protects the BIC, reducing its $Q$-factor to $\sim 10^2$, shown in Fig.\ S6 in the Supplemental Materials. Together, these experimental measurements show that this system possesses a symmetry-protected BIC due to the presence of the surrounding woodpile photonic crystal environment.

Finally, to demonstrate that our experimental system exhibits a line of symmetry-protected BICs, we repeat the experiment for different values of $\theta$, which for $\phi = 0$ correspond to different wavevectors in the Brillouin zone along the $\bold{\Gamma}$--$\bold{X}$ line. As is shown in Fig.\ \ref{fig:3}a, in all cases the resonance is clearly seen for large values of $|\phi|$, but vanishes as $|\phi| \rightarrow 0$. We estimate the frequencies of the BICs in these measurements (see Methods) and find excellent agreement with the frequencies obtained by simulations with an average error between the two of $\Delta \omega/\omega = 1.84\% $, Fig.\ \ref{fig:3}b.


\section{Discussion}
In conclusion, using two-photon polymerization-based 3D-micro printing of a photonic device, we have experimentally demonstrated that three-dimensional photonic crystal environments can be used to create symmetry bandgaps and to realize symmetry-protected BICs away from normal-incidence, which we analytically prove is impossible with homogeneous, isotropic environments. This greatly expands the design space and the achievable properties of BIC-based devices. Moreover, we have shown that not only can symmetry bandgaps be realized in low-index systems, but also that the benefits of environmental design require only a single layer of the environment on both sides of the system. These ideas have immediate ramifications in enabling multi-frequency and multi-wavevector applications in technologies using BICs, such as sensors \cite{yanik_seeing_2011,zhen_enabling_2013,romano_label-free_2018} and high-power lasers \cite{hirose_watt-class_2014}.

\section{Materials and Methods}

\subsection{Fabrication \label{sec:fab}}

Samples are fabricated in the IP-Dip resist (for refractive index in the infrared see \cite{Fullager:17}) using a Nanoscribe Professional GT at a scanspeed of \SI{20}{mm/s} and laser power of 60\% (which corresponds to approximately $\SI{33}{mW}$ on the entrance lens of the objective). The structures were printed onto Menzel cover slips (borosilicate glass) coated with approximately \SI{13}{nm} of $\mathrm{Al}_2\mathrm{O}_3$ to facilitate interface finding in the dip-in configuration. The cover slips have a transmission of greater than 75\% for all wavelengths used in our measurements. After printing, the sample is developed for \SI{10}{min} in PGMEA and \SI{10}{min} in isopropanol, subsequently. In the last step, the sample is blow-dried in a stream of nitrogen. The complete footprint of the structure is approximately \SI{1}{mm^2}. To achieve such a large footprint within reasonable writing time we stitch the structure out of 4x4 angled blocks using stage stitching for larger travel distance. Inside each block and layer we apply piezo-stitching in combination with galvo-scanning for reduced vignetting and more precise positioning. Alignment of stage, piezo and galvo axes is ensured, using the transformation implemented in NanoWrite. Structural parameters are determined via scanning electron microscopy.

For the structure with unbroken symmetry they are as follows: lattice constant $a=\SI{1.01\pm0.01}{\um}$, rod width in $x$, $r_x=\SI{0.50\pm0.04}{\um}$, rod width in $y$, $r_y=\SI{0.56\pm0.06}{\um}$, width of the grating $r_\mathrm{g}=\SI{0.78\pm0.03}{\um}$. The layer heights in $z$ from top to bottom are $h_\mathrm{top,grating}=\SI{1.44\pm0.04}{\um}$, $h_{\mathrm{top,rods,}y}=\SI{1.33\pm0.02}{\um}$, $h_{\mathrm{top, rods,}x}=\SI{1.37\pm0.02}{\um}$, slab height $h_\mathrm{s}=\SI{1.01\pm0.05}{\um}$, $h_{\mathrm{bottom, rods,}x}=\SI{1.45\pm0.05}{\um}$, and $h_{\mathrm{bottom, rods,}y}=\SI{1.45\pm0.02}{\um}$. The bottom grating differs in height across the footprint of the structure due to a slight tilt of the substrate during the fabrication and lies between $\SI{4}{\um}$ and $\SI{5.5}{\um}$. The height of the lower grating was chosen such that any (unpredictable) tilt of the substrate during fabrication would not result in a complete vanishing of the grating anywhere across the structure.

For the symmetry-broken structure (shown in the Supplemental Materials) the parameters are as follows: $a=\SI{1.00\pm0.02}{\um}$,  $r_x=\SI{0.49\pm0.02}{\um}$,  $r_y=\SI{0.59\pm0.01}{\um}$,  $r_\mathrm{g}=\SI{0.68\pm0.01}{\um}$. The layer heights in $z$ from top to bottom are $h_\mathrm{top,grating}=\SI{1.38\pm0.07}{\um}$, $h_{\mathrm{top,rods,}y}=\SI{0.80\pm0.06}{\um}$, $h_{\mathrm{top, rods,}x}=\SI{1.35\pm0.04}{\um}$, slab height $h_\mathrm{s}=\SI{0.82\pm0.04}{\um}$, $h_{\mathrm{bottom, rods,}x}=\SI{1.23\pm0.05}{\um}$, $h_{\mathrm{bottom, rods,}y}=\SI{1.15\pm0.04}{\um}$. The bottom grating differs in height across the footprint of the structure and lies between $\SI{1.1}{\um}$ and $\SI{3.9}{\um}$.

\subsection{Measurement}

To measure the spectra of the samples we use the Hyperion 3000 microscope attached to a Bruker Vertex v70 FTIR. The spectra are taken with a nitrogen cooled MCT detector and a halogen lamp in transmission mode. To increase $k$-space resolution the lower 15x Cassegrain objective is covered except for a pinhole of \SI{1}{mm} in diameter, such that we obtain a nearly collimated beam. We fix the value of $k_x$ (angle $\theta$ to the $x$-axis) by first tilting the sample by $\theta$ around the $y$-axis, and then scan through $k_y$ by tilting the sample with respect to the beam around the $x$-axis. This is done by tilting the sample holder in steps of approximately 0.5°. Since we do not know the exact position of perpendicular incidence of the beam relative to the sample from the setup, we determine $k_y=0$ from the symmetry of the measured angle resolved transmission spectra. All spectra are referenced to the transmission of the used substrates. For each spectrum taken we average over 64 measurements with an FTIR resolution set to \SI{4}{cm^{-1}} in wavenumber.

The small dip in transmission around \SI{3500}{cm^{-1}}, constant across angles, is due to the absorption in the IP-Dip resist \cite{Fullager:17}.

It is difficult to extract the exact frequencies of the BICs from the experiments because the Fano feature of the slab resonance vanishes around the position of the BICs (i.e. the transmission becomes unity). Therefore, the frequencies of the BICs for a given $\theta$ shown in Fig.\ \ref{fig:3}b were estimated the following way: For each $\theta$ we set the upper limit of the BIC frequency to be the horizontal connection between the frequencies for which the resonance was just visible anymore. To obtain the lower limit, we drew a linear continuation of the resonance at both sides of $\phi=0$ and extracted the frequency at their intersection at $\phi=0$. The plotted points in Fig.\ \ref{fig:3}b are then the average of these two frequency limits for each respective $\theta$, while their difference gives the vertical error bar.

\subsection{Numerical methods}

The numerical simulations shown here were performed using three different software packages, \textsc{MIT Electromagnetic Equation Propagation} (MEEP) \cite{oskooi_meep:_2010}, \textsc{MIT Photonic Bands} (MPB) \cite{johnson_block-iterative_2001}, and \textsc{Stanford Stratified Structure Solver} ($S^4$) \cite{S4}. The projected-in-$k_z$ band structures of the rectangular woodpile environment shown in Fig.\ \ref{fig:1}b, as well as the modal profiles of the environment in Fig.\ \ref{fig:1}c were calculated using MPB. Calculations of the quality factor and field profiles of the resonances of the slab, such as the purple line in Fig.\ \ref{fig:1}b, the resonance profile in Fig.\ \ref{fig:1}e, and the properties of the symmetry-detuned system in Fig.\ \ref{fig:1}f, were performed using MEEP. Finally, numerical simulations of the transmission spectrum in Figs.\ \ref{fig:2}d-e were performed using $S^4$. Note, that the purple squares in Figs.\ \ref{fig:2}d-f and the purple data in Fig.\ \ref{fig:3}b were calculated using MEEP.

In simulations of the system with an infinite environment, i.e.\ those in Fig.\ \ref{fig:1} as well as Fig.\ S4, the environmental layers above and below the slab were all taken to have the dimensions shown in Fig.\ \ref{fig:1}a, and the dielectric of all of the photoresist structures is assumed to be $\varepsilon = 2.34$. For Fig.\ S4, the top and bottom period-doubled gratings are taken to both have height $1.4a$ and width $0.79a$. However, simulations of the finite system shown in Figs.\ \ref{fig:2} and \ref{fig:3} all use the scanning electron microscope measured dimensions of the system given in the Fabrication section, and assume that all of the photoresist structures have dielectric $\varepsilon = 2.34 + 0.005i$, to approximate the effects of surface roughness. Note, due to the fabrication process, these layers all overlap. As such, the heights quoted are for the height of each rod from the bottom of that rod to the top, but the vertical separation between adjacent layers of rods is approximated to $a$. In other words, the rectangular woodpile unit cell has dimensions $a \times a \times 2a$. For the $S^4$ simulations in Fig.\ \ref{fig:2}e, a glass substrate, $\varepsilon = 2.25$, with a height of $170a$ was included and the samples are measured (in the experiment, and thus also in the simulations) ``bottom-up,'' i.e., the incident light first transmits or reflects off of the glass substrate, then the ``bottom'' period-doubled grating, and so on, until finally escaping out the other side from the ``top'' grating. Moreover, for $S^4$ simulations in Fig.\ \ref{fig:2}e, the high-frequency Fabry-Perot resonances from the thick glass substrate were filtered out using a Fourier analysis.

\section{Supplementary Materials}

Supplementary material for this article is available at ???

\textbf{This PDF file includes:}

Supplementary Text

Figs.\ S1 to S6

References \cite{inui,taghizadeh_quasi_2017}


\section*{Acknowledgements}
  Computations for this research were performed on the Pennsylvania State University’s Institute for Computational and Data Sciences’ Roar supercomputer. A.C.\ acknowledges support from the Center for Integrated Nanotechnologies, an Office of Science User Facility operated for the U.S.\ Department of Energy (DOE) Office of Science at Sandia National Laboratories. Sandia National Laboratories is a multimission laboratory managed and operated by
  National Technology \& Engineering Solutions of Sandia, LLC, a wholly owned subsidiary of Honeywell International, Inc., for the U.S.\
  DOE’s National Nuclear Security Administration under contract DE-NA-0003525. The views expressed in the article do not necessarily
  represent the views of the U.S.\ DOE or the United States Government.

\section*{Funding}
  A.C.\ acknowledges support from the Laboratory Directed Research and Development program at Sandia National Laboratories. 
  C.J.\ gratefully acknowledges funding from the Alexander von Humboldt Foundation within the Feodor-Lynen Fellowship program. G.v.F.\ acknowledges funding by the Deutsche Forschungsgemeinschaft through CRC/Transregio 185 OSCAR (project No.\ 277625399). W.A.B.\ is grateful for the support of the Eberly Postdoctoral fellowship at the Pennsylvania State University.
  M.C.R.\ acknowledges the support of the U.S.\ Office of Naval Research (ONR) Multidisciplinary University Research Initiative (MURI) under Grant No.\ N00014-20-1-2325 as well as the Charles E. Kaufman foundation under award number KA2020-114794.

\section*{Author contributions}
A.C.\ conceived of the project idea. A.C.\ and C.J.\ designed the device. C.J. fabricated the structures and, assisted by S.A., performed the experimental observation. A.C., assisted by S.V., W.A.B., and C.W.H., performed the theoretical analysis and numerical simulations. All authors discussed the results and contributed to writing the final version of the paper. M.C.R.\ and G.v.F.\ supervised the project. 

\section*{Competing interests}
The authors declare that they have no competing interests.

\section*{Data and code availability}
All data needed to evaluate the conclusions in the paper are present in the paper and/or the Supplementary Materials.

\bibliography{bic_references}

\end{document}


\title{Supplemental Material for: Observation of bound states in the continuum embedded in symmetry bandgaps}
\author{Alexander Cerjan}
\thanks{These two authors contributed equally.}
\affiliation{Department of Physics, The Pennsylvania State University, University Park, Pennsylvania 16802, USA}
\affiliation{Sandia National Laboratories, Albuquerque, New Mexico 87185, USA}
\affiliation{Center for Integrated Nanotechnologies, Sandia National Laboratories, Albuquerque, New Mexico 87185, USA}
\author{Christina J{\" o}rg}
\thanks{These two authors contributed equally.}
\affiliation{Department of Physics, The Pennsylvania State University, University Park, Pennsylvania 16802, USA}
\author{Sachin Vaidya}
\affiliation{Department of Physics, The Pennsylvania State University, University Park, Pennsylvania 16802, USA}
\author{Shyam Augustine}
\affiliation{Physics Department and Research Center OPTIMAS, University of Kaiserslautern, D-67663 Kaiserslautern, Germany}
\author{Wladimir A.\ Benalcazar}
\affiliation{Department of Physics, The Pennsylvania State University, University Park, Pennsylvania 16802, USA}
\author{Chia Wei Hsu}
\affiliation{Ming Hsieh Department of Electrical Engineering, University of Southern California, Los Angeles, California 90089, USA}
\author{Georg von Freymann}
\affiliation{Physics Department and Research Center OPTIMAS, University of Kaiserslautern, D-67663 Kaiserslautern, Germany}
\affiliation{Fraunhofer Institute for Industrial Mathematics ITWM, 67663, Kaiserslautern, Germany}
\author{Mikael C. Rechtsman}
\affiliation{Department of Physics, The Pennsylvania State University, University Park, Pennsylvania 16802, USA}


\setcounter{equation}{0}
\renewcommand{\theequation}{S{\arabic{equation}}}
\setcounter{figure}{0}
\renewcommand{\thefigure}{S{\arabic{figure}}}
\setcounter{section}{0}
\renewcommand{\thesection}{S{\Roman{section}}}

\tableofcontents

\newpage

\section{Bragg-diffraction orders}

A system consisting of a photonic crystal slab of finite thickness embedded in a homogeneous environment, such as vacuum,
possesses discrete translational symmetry in the two-dimensional plane of the photonic crystal,
\begin{equation}
  \epsilon(\vec{r}) = \epsilon(\vec{r} + l_1\vec{a}_1 + l_2\vec{a}_2), \label{eq:sm1}
\end{equation}
where $\vec{a}_{1,2}$ are the lattice vectors of the photonic crystal slab and $l_{1,2} \in \mathbb{Z}$.
As any resonance of the photonic crystal slab must satisfy Maxwell's wave equation, and Maxwell's wave equation commutes with all of the possible
discrete translational symmetry operations in Eq.\ (\ref{eq:sm1}), the resonances of the slab can be constructed to be eigenfunctions of both the wave equation and the symmetry operators simultaneously.
This means that the resonances of the slab can be categorized based on their in-plane wavevector, $\mathbf{k}_\parallel = (k_x, k_y)$,
which are the eigenvalues of the discrete translation operators and
are uniquely defined up to the reciprocal lattice vectors, $\vec{b}_i \cdot \vec{a}_j = 2\pi \delta_{i,j}$. In other words,
resonances of the slab at $\mathbf{k}_\parallel$ are equivalent to, and have the same frequencies as, those found at
$\vec{k}_\parallel + \sum_{i=1,2} m_i \vec{b}_i$, where $m_i \in \mathbb{Z}$.

However, for a radiative channel to exist in the surrounding homogeneous environment, where there is continuous translational symmetry,
the propagating light must satisfy the dispersion relation in that medium,
\begin{equation}
  \frac{n^2 \omega^2}{c^2} = k_x^2 + k_y^2 + k_z^2,
\end{equation}
where $n$ is the refractive index of the environment and $\omega$ is the light's frequency. In particular, this means that for light in the
slab at a given frequency to radiate, it must possess a frequency and in-plane wavevector that yield a real out-of-plane momentum, $k_z$,
\begin{equation}
  k_z = \sqrt{\frac{n^2 \omega^2}{c^2} - \left(\vec{k}_\parallel + \sum_{i=1,2} m_i \vec{b}_i \right)^2}. \label{eq:diffdef}
\end{equation}
Note, that although resonances of the photonic crystal slab at $\vec{k}_\parallel + \sum_{i=1,2} m_i \vec{b}_i$ are equivalent,
these correspond to inequivalent radiative channels in free-space. 
\Eqref{diffdef} thus defines a series of frequency cutoffs for how many radiative channels are available for a given resonance to radiate to.
\begin{itemize}
\item If $n^2\omega^2/c^2 < \vec{k}_\parallel^2$, i.e. $m_i = 0$, the resonance is below the light line and is a bound state, perfectly confined
  by total internal reflection.
\item If $n^2\omega^2/c^2 > \vec{k}_\parallel^2$, but $n^2\omega^2/c^2 < (\vec{k}_\parallel + \sum_i m_i \vec{b}_i)^2$ for $m_i \ne 0$,
  only a pair of radiative channels are available, both have the same $k_z$, but have orthogonal polarizations.
  In this case, the resonance of the photonic crystal slab is said to be below the first Bragg-diffraction limit.
\item If $n^2\omega^2/c^2 > (\vec{k}_\parallel + \sum_i m_i \vec{b}_i)^2$
for some set of $(m_1,m_2) \ne (0,0)$, there are additional pairs of radiative channels available with distinct wavevectors and the resonance is above the first Bragg-diffraction limit, though
it is below some subsequent Bragg-diffraction limit.
\end{itemize}
These different regions are schematically illustrated in Fig.\ 1 in the main text. 


\section{Proof -- No symmetry-protected BICs above the Bragg-diffraction limit}

In this section, we will provide a detailed proof of the claim that:
\begin{enumerate}
  \item[] \textbf{Theorem} \textit{For a two-dimensional photonic system with finite thickness embedded in a homogeneous three-dimensional environment, symmetry-protected BICs
  can only be found below the first Bragg-diffraction limit at $\bold{\Gamma}$.}
\end{enumerate}

The question of whether or not a symmetry-protected BIC can be found in a photonic crystal slab amounts to the
question, `For every choice of $\vec{k}_\parallel$, do the available radiative channels span all of the possible symmetries
of the slab's resonances at that wavevector?' Or in other words, `At every $\vec{k}_\parallel$, do the radiative channels exhaust the
irreducible representations of the little group, $\mathscr{G}(\vec{k}_\parallel)$?'
If there are irreducible representations present in $\mathscr{G}(\vec{k}_\parallel)$ which are unavailable in the radiative
channels, then any slab resonance obeying that combination of symmetries will necessarily be a BIC at that $\vec{k}_\parallel$.
Thus, to prove that symmetry-protected BICs cannot exist above the Bragg-diffraction limit, we must show that above this limit the radiative channels
always span the irreducible representations of $\mathscr{G}(\vec{k}_\parallel)$, regardless of the specific symmetries of the system.

In two dimensions, there are $17$ possible space groups (i.e.\ the `wallpaper groups') which consist of different combinations
of three classes of symmetry operations, rotations, reflections, and glides, which form the point group of the structure,
along with the translational symmetry of the system.
As such, the in-plane symmetry of the photonic crystal slab must be one of these 17 space groups.
Here, we are omitting any consideration of the symmetry of the system in the perpendicular direction for two
reasons. First, many photonic crystal slabs and metasurfaces are constructed on a substrate with a different refractive index
from the material above the structure, and such systems are not symmetric in the perpendicular direction.
Second, even for suspended structures, or those which use an index-matching solution to restore reflection symmetry
about the $xy$-plane ($\sigma_{z}$), it is not possible to use this out-of-plane reflection symmetry to construct a BIC in the slab.
This is because the total possible symmetries of the photonic crystal slab of a $\sigma_z$ symmetric system are simply the direct product of this single out-of-plane symmetry with the in-plane symmetries, but the available radiative channels above and below the slab are degenerate, so linear combinations of these degenerate channels span all of the available out-of-plane symmetries of the slab.

Here, we adopt the notation from Ref.\ 52 from the main text, in which we represent symmetry operations
in the form $\{ \beta | \vec{B} \}$, such that
\begin{equation}
  \{ \beta | \vec{B} \} \vec{r} = \beta \vec{r} + \vec{B},
\end{equation}
where $\beta$ is a $3 \times 3$ matrix (or for two-dimensional systems $2 \times 2$ matrix) which rotates or reflects
$\vec{r}$, while $\vec{B}$ denotes a possible translation. When operating on a function,
this acts as
\begin{equation}
  \{ \beta | \vec{B} \} \psi(\vec{r}) = \psi(\beta^{-1} \vec{r} - \beta^{-1} \vec{B}). \label{eq:trans}
\end{equation}
In this nomenclature, the identity operation is $\{ \varepsilon | 0 \}$, and we use $\{ C_n | 0 \}$ to denote counter-clockwise rotation
by $2\pi/n$ about the $z$-axis, $\{ \sigma | 0 \}$ to denote a reflection about a specified plane, and $\{ \sigma | \boldsymbol{\tau} \ne 0 \}$
to denote a glide operation.

Finally, we remind the reader that the slab resonances at a given $\vec{k}_\parallel$ may have lower symmetry than the point
group of the photonic crystal slab itself. The symmetries, $\{ \beta | \vec{B} \}$, which the slab resonances at $\vec{k}_\parallel$ must obey
(i.e., those symmetry operations which constitute the little group $\mathscr{G}(\vec{k}_\parallel)$)
are those for which
\begin{equation}
  \beta \vec{k}_\parallel = \vec{k}_\parallel + \sum_{i=1,2} m_i \vec{b}_i,
\end{equation}
for some choice of integers $m_{1,2}$, such that $\beta \vec{k}_\parallel$ is equivalent to $\vec{k}_\parallel$ up to a reciprocal lattice vector.
This equivalence is denoted as
\begin{equation}
  \beta \vec{k}_\parallel \doteq \vec{k}_\parallel.
\end{equation}

\subsection{Basis for radiative channels}

To study the available symmetries of the radiative channels of the environment propagating away from the slab in the $+z$ direction, we choose a basis of
these plane wave channels based on their three distinguishing characteristics, $|n_z, n_d, s/p\rangle$, which correspond to:
\begin{itemize}
\item $n_z \in [0,N_z-1]$ is the Bragg-diffraction order the channel is in. This value is ordered in decreasing $k_z$, and
  is indexed from $0$, such that $n_z = 0$ is below the first Bragg-diffraction limit and corresponds to `specular radiation' with $m_{1,2} = 0$ in \eqref{diffdef}. 
  The upper bound $N_z$ is the number of unique values of real $k_z$ for all $m_{1,2}$ given
  $\omega$ and $\vec{k}_\parallel$ of the slab resonance. \\
  \textit{Note} -- for $\vec{k}_\parallel$ along the boundary of the Brillouin zone,
  there are no frequencies which are both above the light line and below the first Bragg-diffraction limit, as $k_z$
  for $m_{1,2} = 0$ is always degenerate with $k_z$ for $m_{1,2}$ corresponding to the neighboring Brillouin zone, see Fig.\ 1 in the main text. 
  In this case, there is no specular radiation, and $n_z \in [1,N_z-1]$. 
\item $n_d \in [0,N_d-1]$ labels the in-plane wavevector of the channel in the environment, $\vec{k}_\parallel + \sum_{i=1,2} m_i \vec{b}_i$.
  The total number of possible in-plane wavevectors for a given Bragg-diffraction order, $N_d$,
  is equal to the number of unique pairs $(m_1, m_2)$ which yield the same $k_z$ via \eqref{diffdef}. \\
  \textit{Note} -- For $\vec{k}_\parallel \ne \bold{\Gamma}$, or $\vec{k}_\parallel = \bold{\Gamma}$ and $n_z \ne 0$, $N_d$ must be an integer multiple
  of the number of rotational symmetry operators which leave $\vec{k}_\parallel$ invariant. \\
  \textit{Proof of note} -- This follows from the definition of the little group of $\vec{k}_\parallel$. If $\{C_n| 0\} \in \mathscr{G}(\vec{k}_\parallel)$,
  then $C_n \vec{k}_\parallel \doteq \vec{k}_\parallel$. But given the stated assumptions ($\vec{k}_\parallel \ne \bold{\Gamma}$, or $\vec{k}_\parallel = \bold{\Gamma}$ and $n_z \ne 0$),
  then $C_n \vec{k}_\parallel \ne \vec{k}_\parallel$ except for $C_n = C_1$, so there must be some other
  pair $(m_1,m_2)$ which yields an equivalent in-plane wavevector with the same $k_z$,
  as rotation about the $z$ axis cannot change the momentum in the $z$ direction.
  As different rotation operations will result in different in-plane wavevectors, these new pairs
  generated $(m_1,m_2)$ from the rotation operations $C_n \ne C_{n'}$ must be unique.
\item $s$/$p$ denotes the polarization of the channel, where $s$ corresponds to the radiation channel
  whose electric field vector is parallel to the photonic crystal slab. For $n_z = 0$ at $\vec{k}_\parallel = \bold{\Gamma}$,
  this choice is arbitrary, but the two polarizations are chosen to be orthogonal.
\end{itemize}

\begin{figure}[h]
  \centering
  \includegraphics[width=0.55\linewidth]{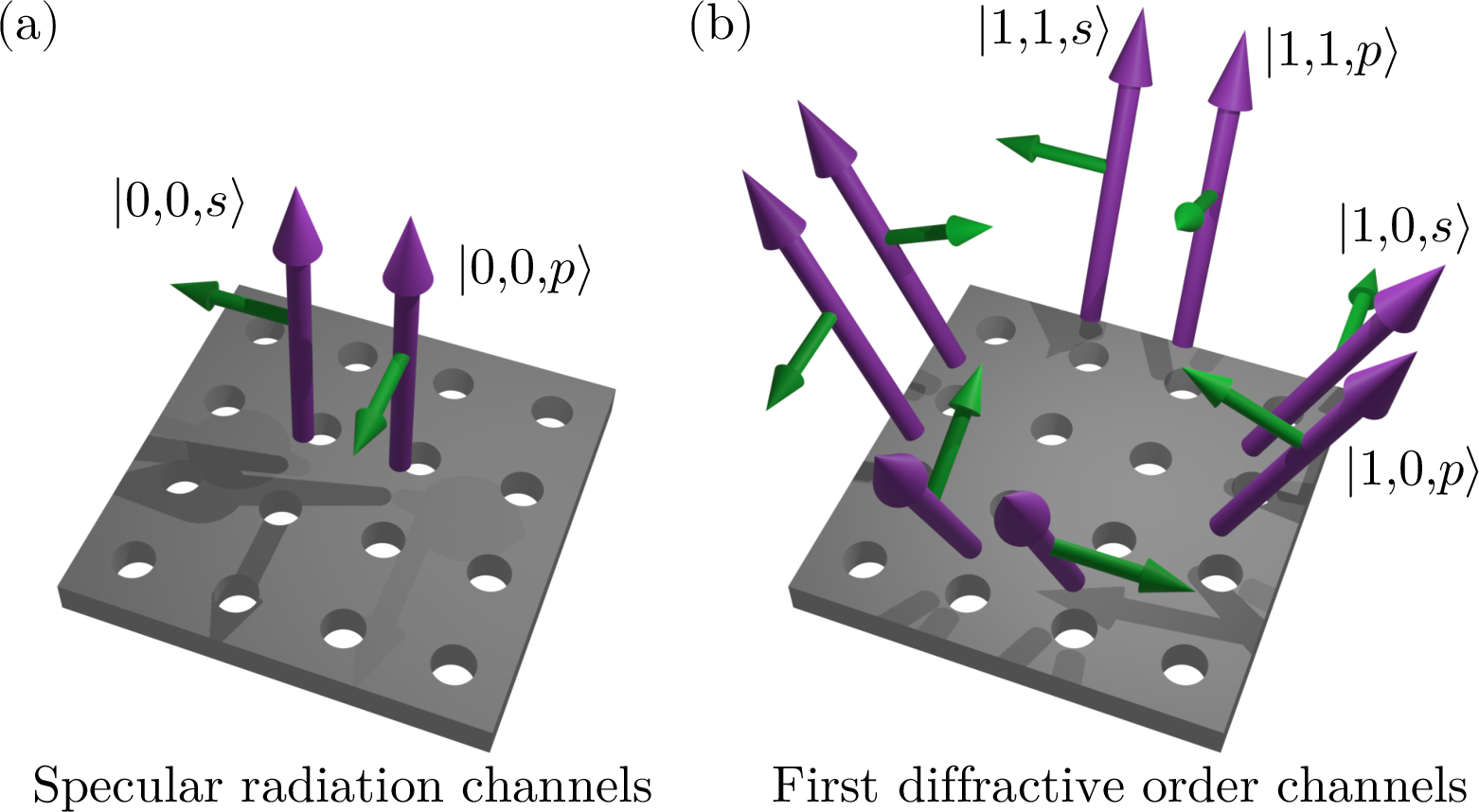}
  \caption{\textbf{Choice of basis for the radiative channels.} Here, the basis, $|n_z, n_d, s/p\rangle$, is shown for a free-space environment at $\vec{k}_\parallel = \bold{\Gamma}$ that surrounds a photonic
    crystal slab whose unit cell is $C_{4v}$ symmetric. (a) Members
    of the basis corresponding to specular radiation, $n_z = 0$. Here, the purple arrow denotes the wavevector of the radiative channel, $\vec{k}$,
    and the green arrow denotes the polarization of the electric field. (b) Members of the basis corresponding to the
    first Bragg-diffraction order, $n_z = 1$. The four unique choices of $(m_1, m_2)$ which constitute this set of in-plane
    wavevectors correspond to $(\pm 1, 0)$ and $(0, \pm 1)$.
  }
  \label{fig:basis}
\end{figure}

An example of this choice of basis for specular radiation, $n_z = 0$ and the first Bragg-diffraction
order, $n_z = 1$, is shown in \figref{basis} for a photonic crystal slab whose unit cell is $C_{4v}$ symmetric at $\vec{k}_\parallel = \bold{\Gamma}$.

Using this basis, it is easy to verify that the representation matrices, $D(\{ \beta | \vec{B} \})$, of the symmetry
operations $\{ \beta | \vec{B} \} \in \mathscr{G}(\vec{k}_\parallel)$ are block-diagonal,
\begin{equation}
  D(\{ \beta | \vec{B} \}) =
  \begin{pmatrix}
    D^{(n_z = 0)}(\{ \beta | \vec{B} \}) & 0 & \cdots \\
    0 & D^{(n_z = 1)}(\{ \beta | \vec{B} \}) & \cdots \\
    \vdots & \vdots & \ddots
  \end{pmatrix},
\end{equation}
in which each block corresponds to the representation of that symmetry operation for a different Bragg-diffraction order of the system. This property is a consequence of the fact
that none of the in-plane symmetry operations can change the out-of-plane wavevector. As such, we can treat the representations
of the different Bragg-diffraction orders separately, and the total symmetries spanned by all of the Bragg-diffraction orders at a given $\vec{k}_\parallel$
will be the direct sum of those symmetries spanned by each order.

\subsection{Structure of the rotation representation matrices}

\begin{figure}[h]
  \centering
  \includegraphics[width=0.30\linewidth]{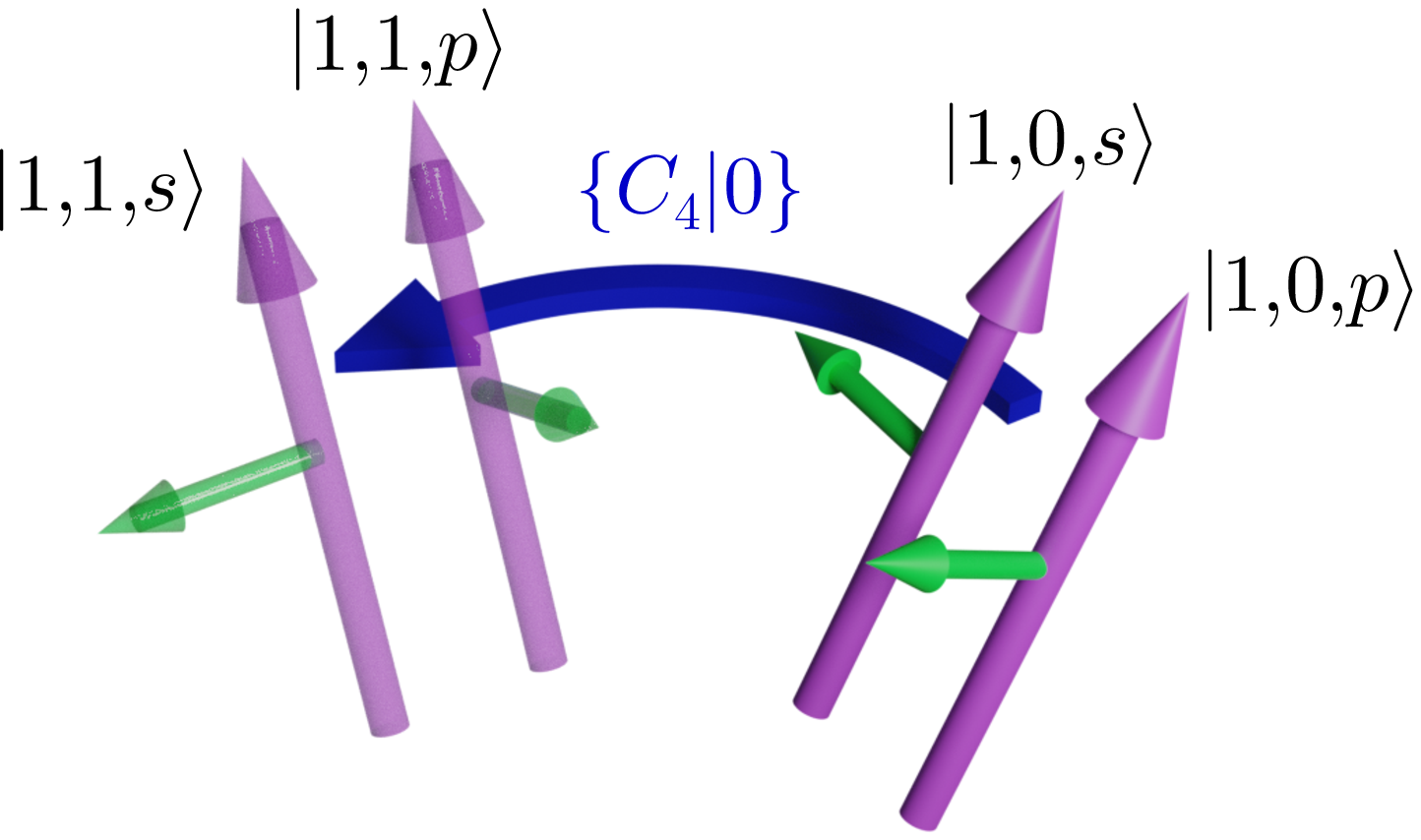}
  \caption{\textbf{Effects of rotation operations.} Example of a $\{C_4 | 0 \}$ rotation operation applied to some basis elements of the system from \figref{basis}.
    For this system,  $D(\{C_4 | 0 \}) |1, 0, s/p\rangle = |1, 1, s/p\rangle$.}
  \label{fig:rot}
\end{figure}

For any non-trivial rotation operation, $\{ C_{n\ne1}| 0\}$, which leaves $C_{n\ne1} \vec{k}_\parallel \doteq \vec{k}_\parallel$,
the effect of this operation on the chosen basis only
results in non-diagonal elements of the representation of this rotation, $D^{(n_z \ge 1)}(\{ C_{n\ne1}| 0\})$, for any non-zero Bragg-diffraction
order. This is a consequence of the fact that for any such radiative channel, $|n_z \ge 1, n_d, s/p\rangle$,
a rotation operation necessarily changes the in-plane component of its wavevector, changing $n_d$. An example of the action
of these rotations on the chosen basis is shown in \figref{rot}.

As such, the character, $\chi(\{ \beta | \vec{B} \}) = \textrm{Tr}[D(\{ \beta | \vec{B} \})]$, of any non-trivial rotation for any non-zero Bragg-diffraction order is zero,
\begin{equation}
  \chi^{(n_z \ge 1)}(\{ C_{n\ne1}| 0\}) = 0.
\end{equation}

\subsection{Structure of the reflection representation matrices}

\begin{figure}[h]
  \centering
  \includegraphics[width=0.55\linewidth]{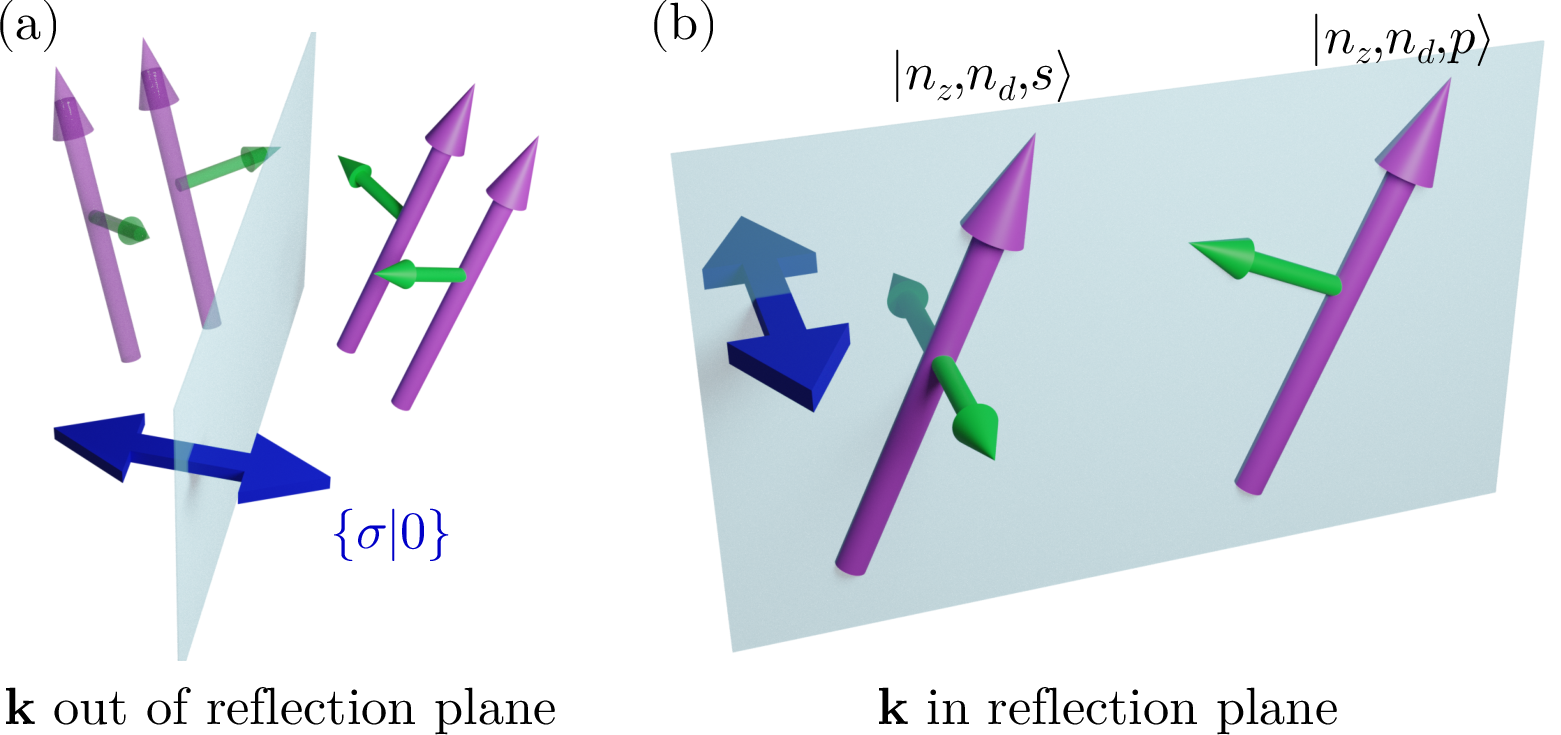}
  \caption{\textbf{Effects of reflection operations.} Examples of reflection operations acting on radiative channel basis functions where the channel's wavevector, $\vec{k}$, either
  lies out of the reflection plane (a), or in the reflection plane (b).}
  \label{fig:refl}
\end{figure}

There are two possible scenarios that can arise when a reflection operation, $\{ \sigma | 0 \}$, which leaves $\sigma \vec{k}_\parallel \doteq \vec{k}_\parallel$,
acts on one of the radiative channel basis functions.
If the basis function's wavevector, $\vec{k}$, does not lie in the reflection plane, then the reflection operation
maps that basis function to a different member of the radiative basis.
However, if $\vec{k}$ lies in the reflection plane then $D(\{ \sigma | 0 \}) |n_z, n_d, s/p\rangle \propto |n_z, n_d, s/p\rangle$.
These two possibilities are shown in \figref{refl}.

Due to the environment being homogeneous, the frequencies and out-of-plane wavevectors for $s$ and $p$ polarized
light are degenerate. Thus, as can be seen in \figref{refl}b, when the wavevector of the radiation channel lies in the reflection
plane,
\begin{eqnarray}
  D(\{ \sigma | 0 \}) |n_z, n_d, s\rangle &=& - |n_z, n_d, s\rangle, \\
  D(\{ \sigma | 0 \}) |n_z, n_d, p\rangle &=& + |n_z, n_d, p\rangle.
\end{eqnarray}
As such, regardless of the particular reflection operation, while the individual diagonal elements of $D(\{ \sigma | 0 \})$ may be non-zero, the character
is necessarily zero,
\begin{equation}
  \chi^{(n_z \ge 1)}(\{ \sigma| 0\}) = 0.
\end{equation}
as $s$ and $p$ polarized channels must appear in the radiative basis in pairs.

\subsection{Structure of the glide representation matrices}

As a reminder, glide operations are the combination of a reflection operation with translation by a vector which cannot be expressed as an integer multiple
of lattice vectors, and can be expressed as the product of these two operations, $\{ \sigma | \boldsymbol{\tau} \} = \{ \varepsilon | \boldsymbol{\tau} \} \{ \sigma | 0 \}$.
Given that we already know the action of the reflection operation on the radiation basis from the previous section,
we must understand the effect of translation on the radiative basis.
As the radiative basis consists of plane wave propagating away from the photonic crystal slab, we know the functional
form of these states far away from the crystal slab,
\begin{equation}
  |n_z, n_d, s/p\rangle = \vec{e}_{s/p} e^{i \vec{k} \cdot \vec{r} - i \omega t},
\end{equation}
where $\vec{e}_{s/p}$ is a unit vector pointing in the direction of the polarization of the electric field.
Thus, using \eqref{trans}, we can calculate the effect of translation on this basis element,
\begin{eqnarray}
  D(\{ \varepsilon | \boldsymbol{\tau} \}) |n_z, n_d, s/p\rangle &=& \vec{e}_{s/p} e^{i \vec{k} \cdot (\vec{r} - \boldsymbol{\tau}) - i \omega t}, \notag \\
  &=& e^{- i \vec{k} \cdot \boldsymbol{\tau}} |n_z, n_d, s/p\rangle.
\end{eqnarray}
As can be seen, this translation effects the two polarizations of radiation channels identically, and thus
again, while the individual diagonal elements of $D(\{ \sigma | \boldsymbol{\tau} \})$ may be non-zero, the character
is necessarily zero,
\begin{equation}
  \chi^{(n_z \ge 1)}(\{ \sigma| \boldsymbol{\tau}\}) = 0,
\end{equation}
regardless of the specifics of the particular glide operation.

\subsection{Completing the proof}

Over the previous subsections, we have proven that the characters for all of the representations of the symmetry operations, $\{ \beta | \vec{B} \}$,
for which $\beta \vec{k}_\parallel \doteq \vec{k}_\parallel$ are zero, except for that of the identity operation,
\begin{eqnarray}
  \chi^{(n_z \ge 1)}(\{ \varepsilon| 0 \}) &=& 2 N_d, \\
  \chi^{(n_z \ge 1)}(\{ \beta | \vec{B} \}) &=& 0, \;\;\;\;\; \textrm{otherwise}.
\end{eqnarray}
Thus, upon using the orthogonality of characters of irreducible representations (see Ref.\ 52) to decompose the representations
of the non-zero Bragg-diffraction orders of the radiative channels, we find that
\begin{eqnarray}
  q_{(\alpha)} &=& \frac{1}{g} \sum_{\{ \beta | \vec{B} \} \in \mathscr{G}(\vec{k}_\parallel)} \chi_{(\alpha)}(\{ \beta | \vec{B} \})^* \chi^{(n_z \ge 1)}(\{ \beta | \vec{B} \}) \notag \\
  &=& \frac{1}{g} \chi_{(\alpha)}(\{ \varepsilon | 0 \})^* \chi^{(n_z \ge 1)}(\{ \varepsilon | 0 \}) \notag \\
  &=& \frac{2N_d}{g} \chi_{(\alpha)}(\{ \varepsilon | 0 \})^*.
\end{eqnarray}
Here, $\alpha$ denotes an irreducible representation of $\mathscr{G}(\vec{k}_\parallel)$, $\chi_{(\alpha)}(\{ \beta | \vec{B} \})$ is the character of that
irreducible representation for $\{ \beta | \vec{B} \} \in \mathscr{G}(\vec{k}_\parallel)$, $g$ is the number of elements in $\mathscr{G}(\vec{k}_\parallel)$,
and $q_{(\alpha)}$ is the number of times that irreducible representation appears in the representation of the environment's radiative channels.
As the character of the identity operation for every irreducible representation is its dimension, $\chi_{(\alpha)}(\{ \varepsilon | 0 \})^* \ge 1$,
and so
\begin{equation}
  q_{(\alpha)} > 0 \;\;\;\;\; \forall \alpha.
\end{equation}

Thus, for any non-zero Bragg-diffraction order of a homogeneous radiative environment surrounding a photonic crystal slab, the radiative channels completely
span the possible symmetries of the modes of the photonic crystal slab. As such, one cannot find symmetry-protected bound states in the continuum
above the Bragg-diffraction limit. In particular, this means that there cannot be symmetry-protected BICs at the high-symmetry points along the boundary
of the Brillouin zone, as the boundary of the Brillouin zone is always above the first Bragg-diffraction limit.

\section{Proof -- No symmetry-protected BICs below Bragg-diffraction limit except at $\vec{k}_\parallel = \bold{\Gamma}$}

For the 17 possible space groups of two-dimensional systems, the only possible symmetry classes which can be found on the interior of the Brillouin zone away
from $\vec{k}_\parallel = \bold{\Gamma}$ is even/odd symmetry with respect to a reflection (or glide) operation. In other words, except for $\bold{\Gamma}$ (and the boundary
of the Brillouin zone which is above the first Bragg-diffraction limit), the maximum number of elements which can be found in $\mathscr{G}(\vec{k}_\parallel)$
is $2$, along high-symmetry lines for which $\vec{k}_\parallel$ lies in the plane of a reflection or glide symmetry. In this case, the point group of $\mathscr{G}(\vec{k}_\parallel)$
is $C_{1h}$, and resonances of the photonic crystal slab at this wavevector will either be even or odd with respect to this reflection or glide symmetry operation.
However, one can quickly verify that the specular radiation channels in the surrounding environment (those with $n_z = 0$), span the two
possible symmetries of the slab's resonances, as the $s$ polarized channel will be odd about the reflection operation, while the
$p$ polarized channel will be even, see \figref{refl}b.

Thus, below the first Bragg-diffraction limit for $\vec{k}_\parallel \ne \bold{\Gamma}$, there are no symmetry-protected BICs, as the
specular radiation channels exhaust the possible symmetries of the resonances of the slab.

\section{Truncation of the environment}

\subsection{Truncation in $z$}
In this section, we numerically justify the truncation of the infinite environment to an environmental layer a single unit cell
thick on each side of the slab. In the absence of an infinite rectangular woodpile environment surrounding the slab, and with the addition of the
period-doubled grating, the symmetry-protected resonance of the slab can evanescently couple to the grating layers and radiate into
the surrounding air, as the resonance along the $\bold{\Gamma}$--$\bold{X}$ line is not a symmetry-protected BIC when the radiative environment is free space.
In this picture, the layers of the rectangular woodpile act as a barrier without any states the slab resonance can couple to, which increases the distance
between the slab and the grating. This lack of states with the correct symmetry in the woodpile photonic crystal means that the coupling between
the slab resonance and the grating is strictly evanescent, and thus decreases exponentially with distance.

\begin{figure}[b]
  \centering
  \includegraphics[width=0.7\linewidth]{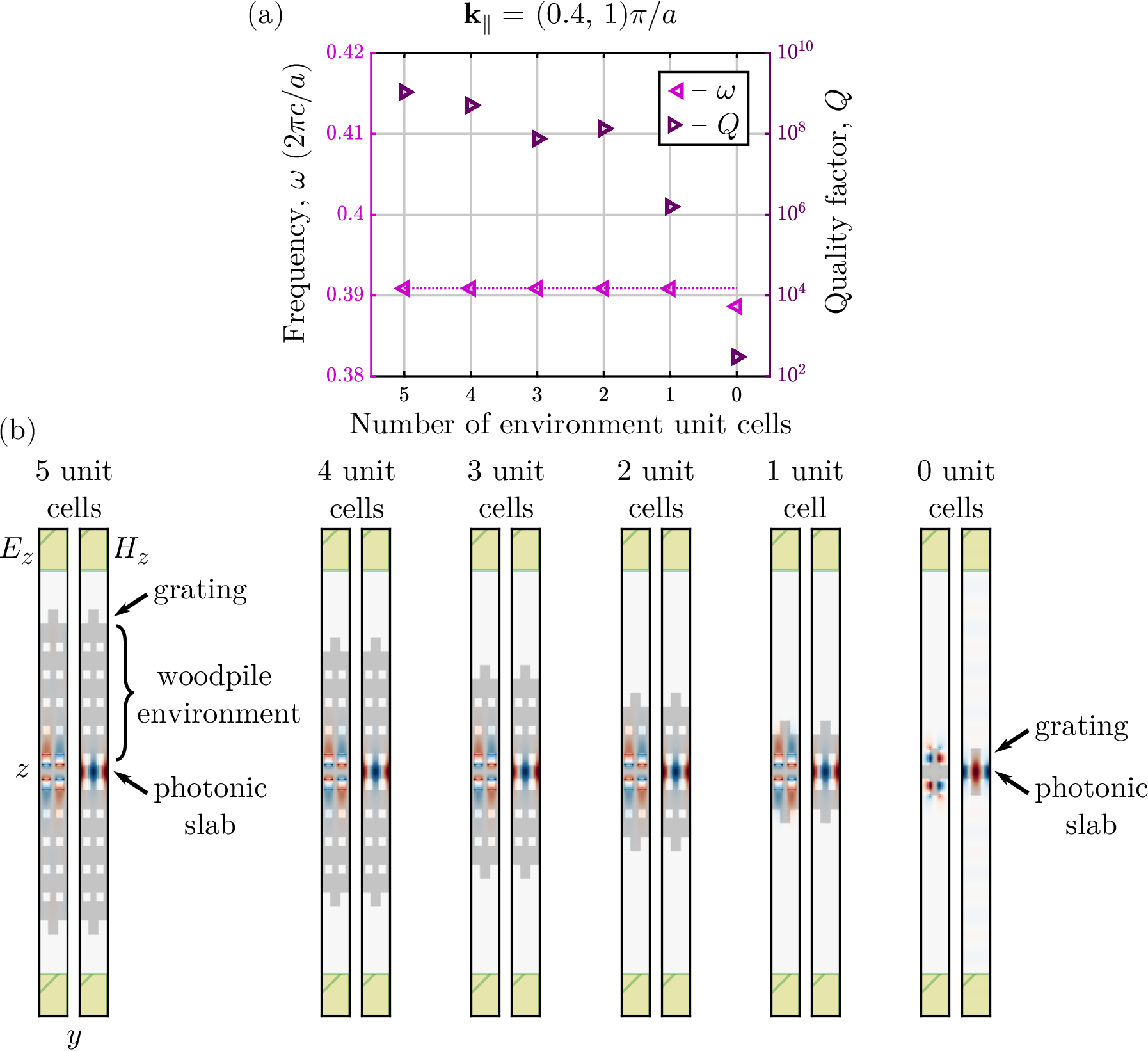}
  \caption{\textbf{Truncation in the out-of-plane direction.} (a) Frequencies (left triangles) and quality factors (right triangles) of the photonic slab resonance as the environment is truncated from
    from being 5 unit cells thick on each side to vanishing. This system is the same as the one considered in Fig.\ 3 of the main text, and includes
    the period-doubled grating.
    (b) Plot of the $E_z$ and $H_z$ field profiles in the $yz$-plane through this truncation process. The gray regions denote the dielectric regions,
    and the yellow boundaries of these plots denote the regions of the perfectly-matched layers absorbing boundary condition.}
  \label{fig:truncation}
\end{figure}

Numerically, we can confirm that the resonance frequency and modal profile remain essentially unchanged when the environment
is truncated from being infinite, to being comprised of $5$ layers of photonic crystal on each side of the slab, see Fig.\ \ref{fig:truncation}. Moreover, the
frequency and modal profile remain nearly fixed as the environment is reduced to being only a single layer of the photonic crystal
on each side, with this state still possessing a $Q$-factor $> 10^6$.
However, both of these features dramatically change if the environment is removed entirely. Then, the slab resonance can couple
directly to the grating, changing its frequency and modal profile, and significantly reducing its $Q$-factor.

Thus, the truncation of the environment to a single unit cell thick on either side of the slab is justified, and does not alter
the essential physics of the system. In contrast, completely removing the environment would significantly change the properties of the slab
resonances.

\subsection{Truncation perpendicular to the slab plane}

Truncation of the system in a direction perpendicular to the slab plane, such as $x$ or $y$, can result in a variety of outcomes depending upon the details of the truncation. For this subsection, we will assume that the environment, where it exists, remains infinite in $z$.

\begin{figure}[t]
  \centering
  \includegraphics[width=0.90\linewidth]{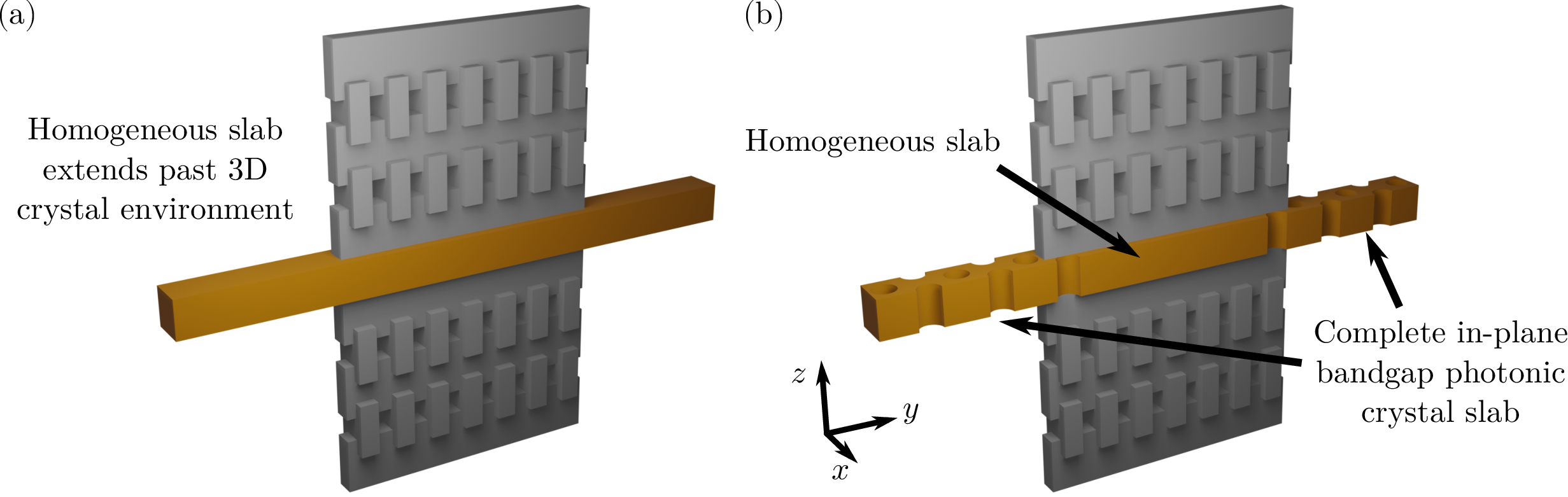}
  \caption{\textbf{Truncation in a direction perpendicular to the slab plane.} These schematics are to be seen as unit cells that are repeated periodically in the $x$-direction. (a) Homogeneous slab extends beyond the truncated environment.
    (b) Homogeneous slab is terminated at a photonic crystal slab. The photonic crystal slab is designed to possess a complete in-plane photonic bandgap at the frequency of the symmetry bandgap.}
  \label{fig:truncation2}
\end{figure}

If only the environment is truncated in a direction perpendicular to the slab plane, but the slab itself continues so that there is an area where the slab is not surrounded by the environment (Fig.\ \ref{fig:truncation2}a), this will yield an effective decay rate to the slab states, turning the BICs into finite-Q resonances. Without the environment, the slab states can couple to the surrounding free-space radiative channels, and thus leak. However, the finite-Q resonances will still essentially form a line, as the slab remains infinite in the in-plane directions.

If the environment is truncated in an in-plane direction \textit{and} the slab is also truncated with an in-plane mirror, say by adding an in-plane photonic crystal slab decoration with a complete bandgap where the environment has been removed (Fig.\ \ref{fig:truncation2}b), then the BICs will become finite-Q but still remain high-Q, as the solid slab is still completely surrounded by the environment, while propagation outside of this region is prohibited by the in-plane mirror at the boundary due to the photonic crystal slab. Then:
\begin{itemize}
\item If the truncation results in a large region of solid slab surrounded by a environment and the line of BICs exists along the remaining infinite in-plane direction, the line of BICs will be nearly preserved. An example such setup is shown in Fig.\ \ref{fig:truncation2}b, where the system retains discrete translational symmetry in $x$ indefinitely (and thus $k_x$ remains a well-defined wave vector component), but the homogeneous slab is finite in $y$. Here, the line of BICs previously along $\bold{Y}$--$\bold{M}$ will remain a line of high-Q resonances if the truncation length in $y$ is sufficiently large such that the standing waves which form in $y$ with the nearly correct symmetry still fall within the original symmetry bandgap of the infinite environment.
\item However, if the in-plane truncation results in only a finite width of slab with the environmental cladding and this finite truncation is in the direction of the line of BICs, this will discretize the line of BICs into a set of high-Q resonances at points with spacing $\delta k = 2\pi/L$, where $L$ is the width of the solid slab in the finite direction. This is similar to the effect discussed in Ref.\ 53 of the main text. An example of such a system is also shown in Fig.\ \ref{fig:truncation2}b, where we now focus on the line of BICs in the non-truncated system along the $\bold{X}$--$\bold{M}$ high symmetry line. When the system is truncated in $y$, $k_y$ is no longer a well-defined wave vector component, so the line of BICs will discretize into a set of points that remain high-Q.
\end{itemize}

\section{Broken Symmetry Experiment}

\begin{figure}[h]
  \centering
  \includegraphics[width=0.4\linewidth]{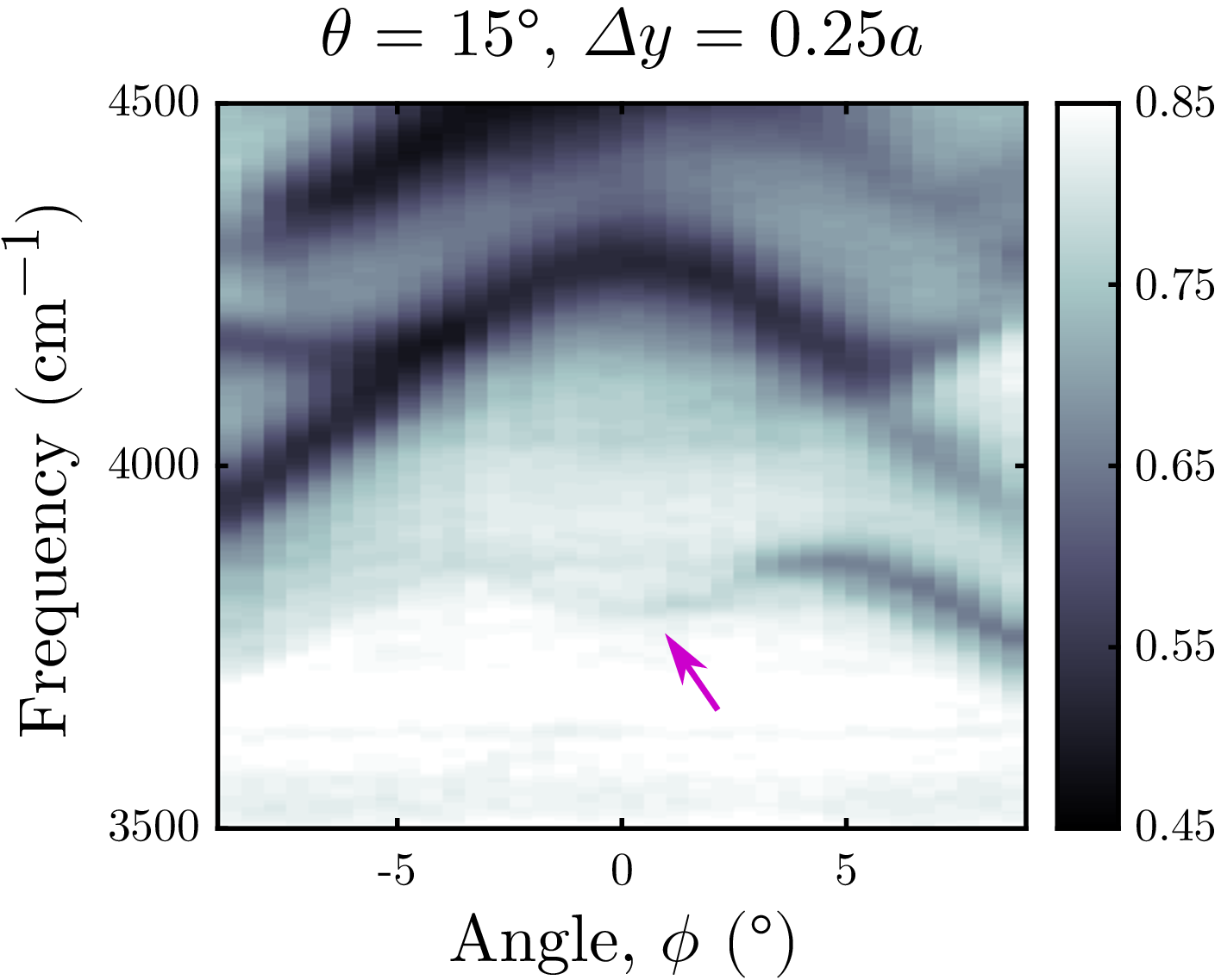}
  \caption{\textbf{Experimentally observed transmission spectrum for a symmetry-broken structure.} Here, we use $\Delta y = 0.25a$ at $\theta = 15^\circ$. See Fig.\ 2f,g in the main text for the definition of $\Delta y$. The purple arrow points to the location of the former-BIC resonance at $\phi = 0^\circ$.}
  \label{fig:broken}
\end{figure}

To experimentally demonstrate that the BIC is protected by the symmetry of the system, we purposefully break this symmetry in the same manner as is shown in Fig.\ 2f,g of the main text. As can be seen in Fig.\ \ref{fig:broken}, for a system with broken symmetry that is otherwise similar to the system used to observe Figs.\ 3e and 4 in the main text (see Materials and Methods section), the resonance now has a finite linewidth at $\phi = 0^\circ$ which can be observed using our experimental methods.

